\documentclass[aps,prl,preprint,groupedaddress]{revtex4}
\usepackage{graphicx}
\begin{document}

\title{Screening and antiscreening in fullerene-like cages: 
dipole-field amplification with ionic nanocages}

\author{Pier Luigi Silvestrelli$^1$}
\email[]{pierluigi.silvestrelli@unipd.it}
\author{S. Subashchandrabose$^{1,2}$}
\author{Abdolvahab Seif$^1$}
\author{Alberto Ambrosetti$^1$}
\affiliation{1) Dipartimento di Fisica e Astronomia, Universit\`a degli Studi di 
Padova, 35131 Padova, Italy}
\affiliation{2) Centre for Research and Development, Department of Physics, PRIST 
Deemed University, Thanjavur, Tamilnadu-613403, India}

\date{\today}

\begin{abstract}
The successful synthesis of endohedral complexes consisting of nanoscale 
carbon cages that can encapsulate small molecules has been a
remarkable accomplishment since these systems are ideal models to investigate
how confinement effects can induce changes in structural and electronic 
properties of encapsulated molecular species.
We here investigate from first principles screening effects observed 
when small molecules, characterized by a finite electronic dipole moment,
such as HF, LiF, NaCl, and H$_2$O, are encapsulated into different nanoscale
cages: C$_{60}$, C$_{72}$, B$_{36}$N$_{36}$, Be$_{36}$O$_{36}$,
Li$_{36}$F$_{36}$, Li$_{36}$Cl$_{36}$, Na$_{36}$F$_{36}$, Na$_{36}$Cl$_{36}$,
and K$_{36}$Br$_{36}$.
%Binding energies, equilibrium geometries, electronic properties,
%and vibrational frequencies of these complexes have been computed.
Binding energies and electronic properties, of these complexes have been computed.
In particular, detailed analysis of the effective dipole moment of the
complexes and of the electronic charge distribution suggests that
screening effects crucially depend on the nature of the
intramolecular bonds of the cage: screening is maximum
in covalent-bond carbon nanocages, while it is reduced in partially-ionic
nanocages B$_{36}$N$_{36}$ and Be$_{36}$O$_{36}$, being very small in the
latter cage which turns out to be almost ``electrically transparent''. 
Interestingly, in the case of the ionic-bond nanocages, an 
antiscreening effect is observed: in fact, due to the relative displacement 
of positive and negative ions, induced by the dipole moment of the 
encapsulated molecule, these cages act as dipole-field amplifiers.
Our results open the way to the possibility of tuning the dipole moment of
nanocages and of generating electrostatic fields at the nanoscale without the 
aid of external potentials. Moreover, we can expect some transferability of the 
observed screening effects also to nanotubes and 2D materials.
\end{abstract}


\maketitle

\section{Introduction}
Buckminsterfullerene (C$_{60}$) is a carbon nanostructured allotrope with
a cage-like fused-ring structure (truncated icosahedron) made of 20 
carbon hexagons and 12 carbon pentagons where each carbon atom has three 
bonds. Since its discovery\cite{Kroto} this complex has 
received intense study, also considering that, although C$_{60}$ is the most
stable and the most common naturally occurring fullerene, many other 
cage-like nanostructures have been obtained and can be hypothesized, 
by both considering different
numbers of C atoms and also replacing carbons with other atoms.
For instance, it has been natural to search for cages made by B and N atoms,
since the B-N pair is isoelectronic with a pair of C atoms; however, a fullerene
structure made by 60 B and N atoms is not optimal since the presence of
pentagonal rings does not allow a complete alternate sequence of B and N atoms.
Fullerene-like alternate B-N cages can be formed introducing isolated squares
characterized by 4 B-N bonds with alternate B and N atoms.
In particular, a structure made by 36 B and 36 N atoms (B$_{36}$N$_{36}$), 
with a relatively large energy gap between the highest molecular 
orbital (HOMO) and the lowest molecular orbital (LUMO), has been found to 
be energetically very stable,
both in theoretical first-principles studies and experimental investigations
(see ref. \onlinecite{Zope} and references therein).

Interestingly, by high-energy collisions of ionized fullerene species, harsh
conditions of high temperature and pressure, electric arc, or by organic
synthesis methods (``molecular surgery''), it is nowadays possible to 
produce C$_{60}$ endohedral complexes with metal ions, noble gases,
and small molecules, such as H$_2$, N$_2$, H$_2$O, and CH$_4$ 
(the first organic molecule to be 
encapsulated)\cite{Akasaka,Suetsuna,Komatsu,Kurotobi,Bloodworth,Jaworski}.
Such recent achievements in the synthesis of endohedral fullerene complexes
have stimulated many experimental and theoretical investigations since 
the cavity inside fullerenes provides a unique environment for the study of 
isolated atoms and molecules. Moreover, these systems represent ideal models 
to study how confinement effects can induce changes in structural and 
electronic properties of small molecular species and also provide a 
possible way to alter the properties of the otherwise rather inert fullerenes.
In particular, Kurotobi and Murata developed a synthetic route to surgically 
insert a single water molecule into the most common fullerene 
C$_{60}$\cite{Kurotobi}, a remarkable achievement considering
that water under normal conditions prefers to exist in a hydrogen bond
forming hydrophilic environment. The water molecule, with its relatively
large dipole moment (1.9 D), is expected to polarize the 
symmetric non-polar C$_{60}$ cage. However,
the theoretical study of such polarization effects has given rise to a 
scientific controversy. In fact, while Kurotobi and Murata\cite{Kurotobi}, 
and Bucher\cite{Bucher}
estimated a surprisingly high value of the dipole moment of the H$_2$O@C$_{60}$
complex (a value similar to that of the isolated water molecule),
other theoretical first-principles studies\cite{Ramachandran,Yagi,Ensing}
indicate that the dipole moment of H$_2$O@C$_{60}$ is instead
much lower (about 0.5 D) than that for the isolated water, thus suggesting
that a substantial counteracting dipole moment is induced in the 
C$_{60}$ cage, which considerably screens the electric field produced by 
the dipole moment of the encapsulated water molecule. 
The residual dipole moment of H$_2$O@C$_{60}$ is still significant, which could
have interesting implications for possible applications of fullerenes.

In this work, by adopting independent theoretical approaches, we 
confirm our previous conclusions\cite{Ensing} about the pronounced 
screening of the dipole moment of a water molecule encapsulated into
C$_{60}$ and extend the study to the encapsulation of some linear diatomic
molecules, characterized by a dipole moment comparable (HF) to or even
much larger (LiF and NaCl) than that of water.  
We also investigate screening effects in other 
cage-like nanostructures, such as B$_{36}$N$_{36}$, Be$_{36}$O$_{36}$, 
C$_{72}$ (a carbon fullerene with the same structure of B$_{36}$N$_{36}$), 
and the hypothetical ionic-bond cages (again with the 
same structure of B$_{36}$N$_{36}$) Li$_{36}$F$_{36}$, Li$_{36}$Cl$_{36}$, 
Na$_{36}$F$_{36}$, Na$_{36}$Cl$_{36}$, and K$_{36}$Br$_{36}$.

In Figs. 1-3 we show some of the investigated nanocages,
namely C$_{72}$, B$_{36}$N$_{36}$, and Li$_{36}$F$_{36}$, characterized by
{\it covalent}-bonds, {\it partially-ionic} bonds, and 
{\it predominantly-ionic} bond,
respectively (the figures are also representative of the other 
considered systems).
We also plot the electron charge distribution to highlight the different
bonding character of the nanocages.

\begin{figure}
%{\vskip 1.3cm}
\centerline{
\includegraphics[width=8cm]{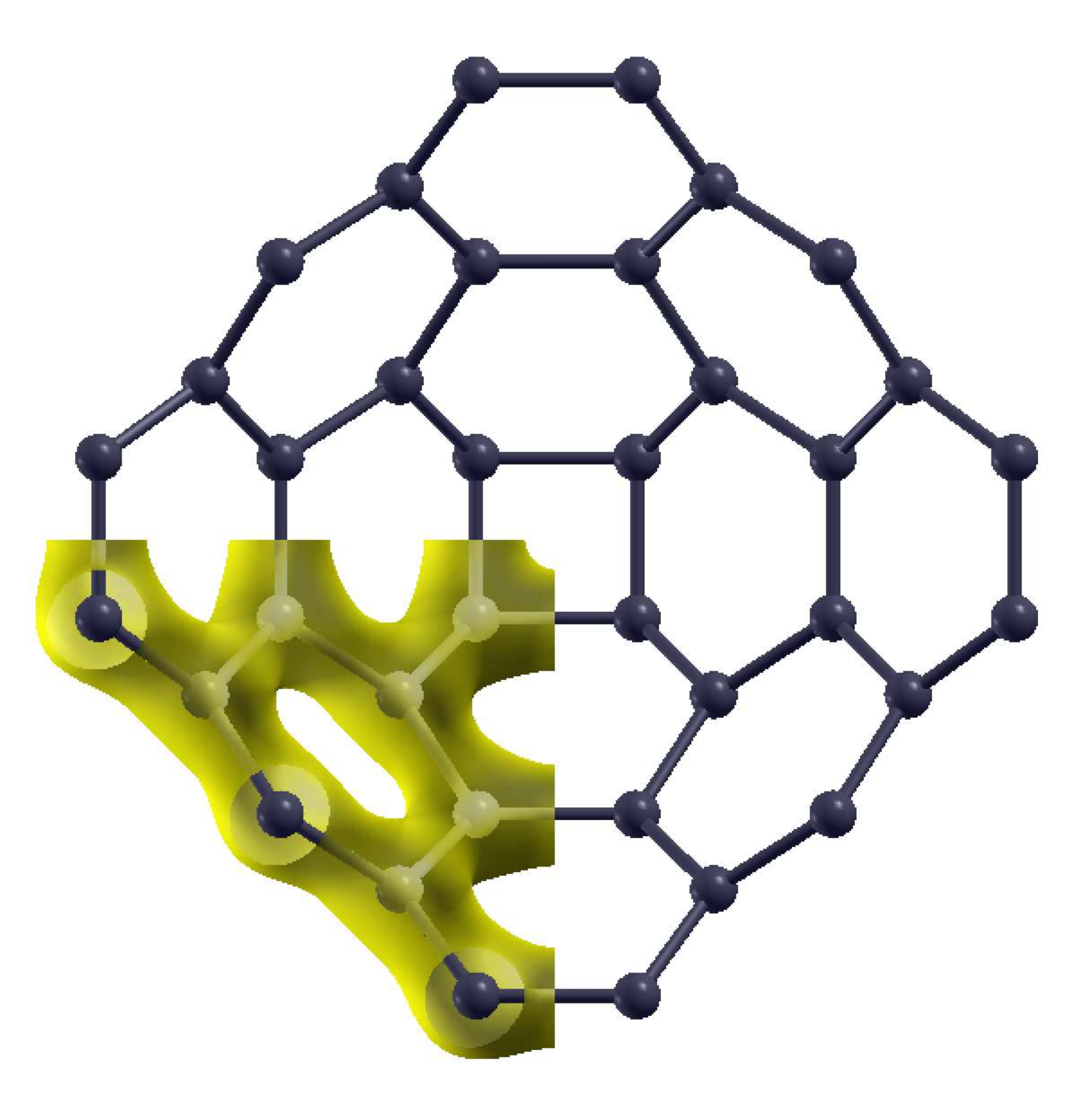}
}
\caption{C$_{72}$ nanocage. The electron charge distribution corresponding
to an isosurface of 1.0 $e/{\rm \AA}^3$ is also partially
plotted (in a quarter of the figure).}
\label{fig1}
%\huge
\end{figure}
\eject
                      
\begin{figure}
%{\vskip 1.3cm}
\centerline{
\includegraphics[width=8cm]{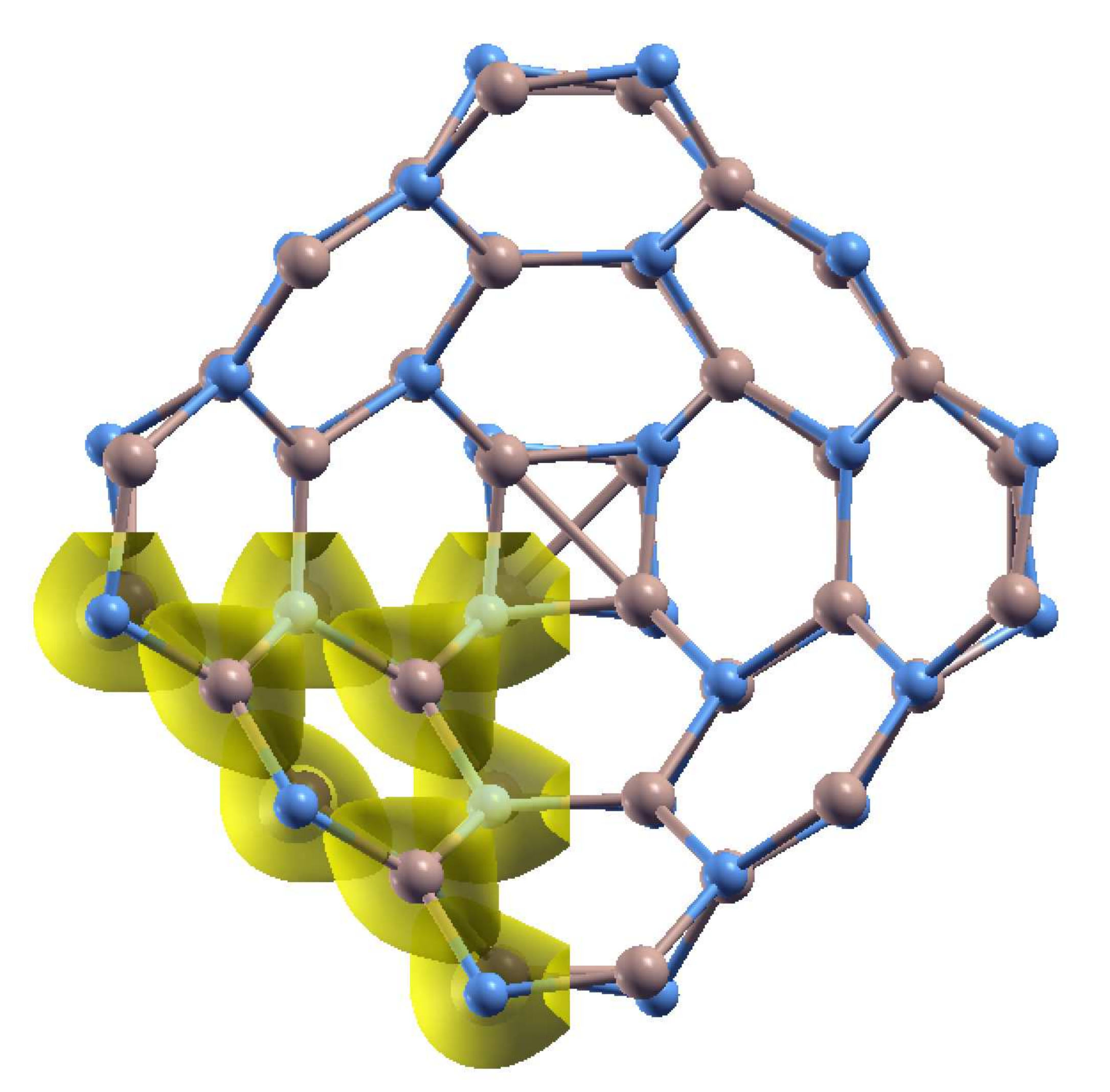}
}
\caption{B$_{36}$N$_{36}$ nanocage. Brown and blue balls represent B and N atoms,
respectively.The electron charge distribution corresponding
to an isosurface of 1.0 $e/{\rm \AA}^3$ is also partially
plotted (in a quarter of the figure).}
\label{fig2}
%\huge
\end{figure}
\eject
                      
\begin{figure}
%{\vskip 1.3cm}
\centerline{
\includegraphics[width=8cm]{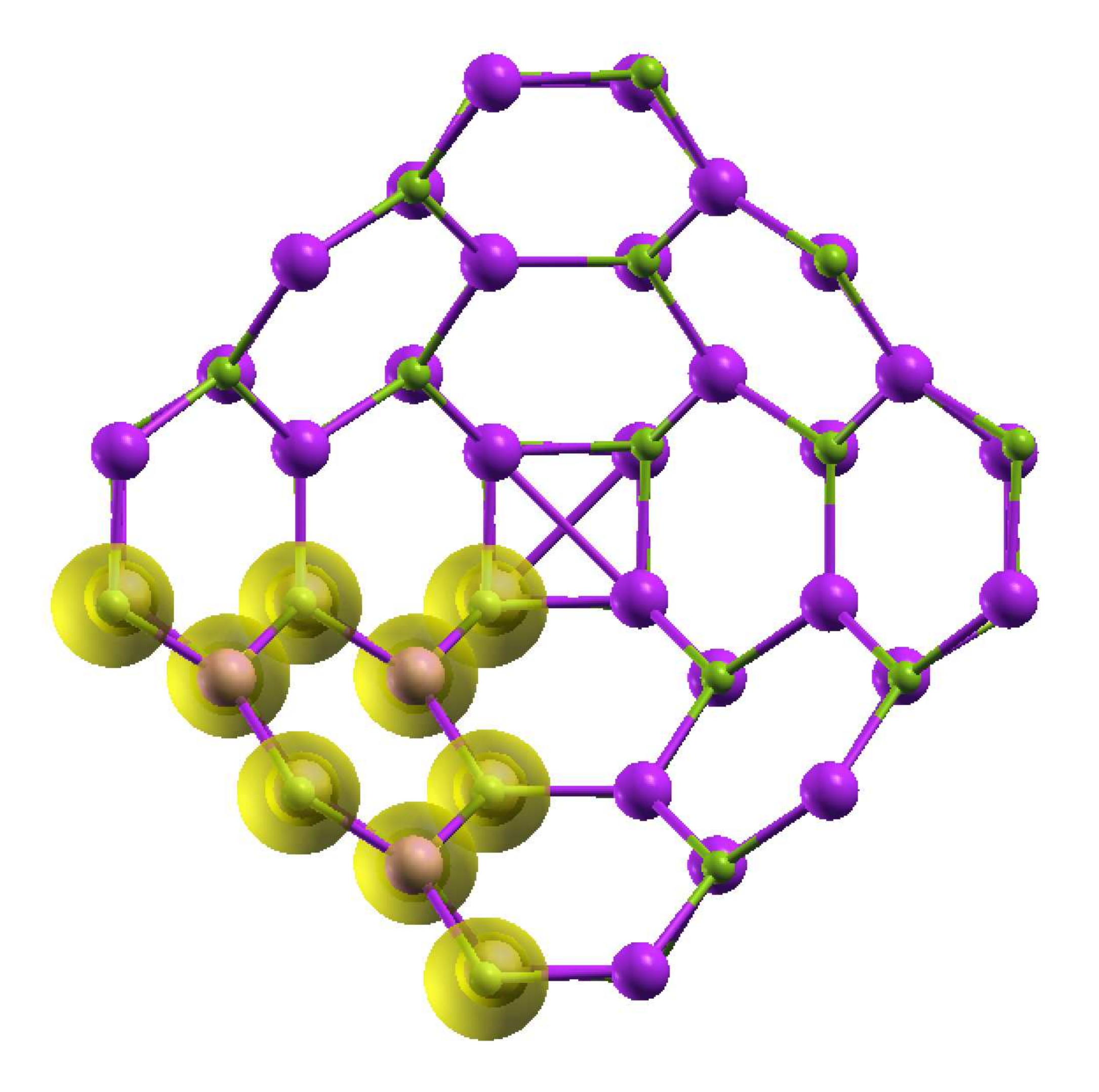}
}
\caption{Li$_{36}$F$_{36}$ nanocage. Violet and green balls represent Li and F atoms,
respectively. The electron charge distribution corresponding
to an isosurface of 1.0 $e/{\rm \AA}^3$ is also partially
plotted (in a quarter of the figure).}
\label{fig3}
%\huge
\end{figure}
\eject
                      
%Basically, our calculations of binding, structural, vibrational 
Basically, our calculations of binding and
electronic properties, and detailed analysis of the effective 
dipole moment of the
complexes and the electronic charge distribution elucidate
the encapsulation effects and suggest that
the screening phenomenon crucially depends on the nature of the
intramolecular bonds of the cage: screening is maximum
in covalent-bond carbon nanocages, is reduced in partially-ionic
ones, while in the case of the ionic-bond nanocages, an 
{\it antiscreening} effect is observed. Hence, the latter systems surprisingly act 
as dipole-field amplifiers.

\section{Methods}
Our first-principles simulations have been performed
with the Quantum-ESPRESSO {\it ab initio} package\cite{ESPRESSO}, 
within the framework of the Density Functional Theory (DFT).
The investigated systems are located in periodically repeated 
cubic supercells, sufficiently large (finite-size effects have been
carefully tested) to avoid significant spurious interactions due to periodic
replicas: the lattice side ranges from 30 to 40 a.u., depending on the
nanocage diameter. 
As a consequence, the sampling of the Brillouin Zone has been
restricted to the $\Gamma$-point only.
Electron-ion interactions were described using ultrasoft
pseudopotentials and the wavefunctions were expanded in a plane-wave basis
set with an energy cutoff ranging from 34 to 80 Ry, depending on the 
atomic elements of the system.
Since van der Waals (vdW) forces are expected to play an important role 
in the interaction of an encapsulated molecule with the surrounding 
cage\cite{Ensing},
the calculations have been performed by adopting the
rVV10 DFT functional\cite{Sabatini} (this is the revised, more efficient
version of the original VV10 scheme\cite{Vydrov}), where
vdW effects are included by introducing an explicitly nonlocal correlation
functional.
rVV10 has been found to perform well in many systems and phenomena
where vdW effects are relevant, including several adsorption
processes\cite{Sabatini,psil15,psil16}.

In order to corroborate the conclusions of our DFT-rVV10 calculations
and better elucidate screening effects,
we have also studied some of our systems by an alternative approach, namely
the Self-Consistent Screening scheme (SCS)\cite{Ambrosetti2014}. 
The SCS approach maps the dipole polarizability of the system into a set of
coupled atom-centered Drude oscillators. The oscillators are parametrized 
according to the Tkatchenko-Scheffler\cite{ts} approach
to account for charge hybridization. Moreover, SCS includes long-range many-body contributions 
up to infinite order, at an effective\cite{Tkatchenko} 
Random Phase Approximation (RPA) level. In practice, one solves a discrete 
self-consistent electrostatic equation, 
that describes the coupled atomic polarizabilities in the presence of an external field.  
We also note that since SCS relies on atomic polarizabilities, it is a {\it linear} 
response theory by construction.

\section{Results}
In Table I we report some basic properties of the cage-like nanostructures
considered in the present study: the cage diameter, the cohesive energy, and
the HOMO-LUMO energy gap $E_g$. For all the nanocages consisting of two 
different types of atoms $E_g$ is considerable (ranging from 4.0 to 5.9 eV)
and significantly larger than that of the carbon nanocages. 
The cohesive energy (per atom) is defined as:
\begin{equation}
E_c = ( E - \sum_i^{N} E_i )/N  \;,
\end{equation}
where $E$ is the total energy of the system, $E_i$ is the energy of the
isolated $i$-th atom, and $N$ is the total number of
atoms of the system.
As can be seen, the cohesive energy is comparable for C$_{60}$, C$_{72}$,
and B$_{36}$N$_{36}$, it is slightly smaller (in absolute value) for
Be$_{36}$O$_{36}$, while it is considerably smaller for the ionic nanocages,
although, for instance, $E_c$ of Li$_{36}$F$_{36}$ is significantly larger 
than that of the Si$_{60}$ cage, which was found to be structurally stable
at room temperature; actually Si$_{60}$ turns out to be stable towards
spontaneous disintegration up to 700 K, according to previous
first-principles simulations\cite{Chen}.
Also note that first-principles calculations showed 
the high stability of alkali-halide LiF nanotubes\cite{Lima} and stable nanotube
structures were also found for NaCl and KBr.
E$_c$ of the other considered ionic cages is instead smaller,
so that we expect that these structures are less stable than
Li$_{36}$F$_{36}$.
%In order to directly assess the stability of the ionic 
%nanocages we have carried out Molecular
In order to directly assess the stability of the Li$_{36}$F$_{36}$ 
nanocage we have carried out Molecular
Dynamics (MD) simulations at different average temperatures.
Newton's equations were integrated using the Verlet algorithm with a
MD time step of 1.0 fs and typical total simulation times of 1 ps; the 
%MD time step of 0.97 fs and typical total simulation times of 1 ps; the 
ionic temperature was set to the desired value by simple velocity rescaling.
We observe that at T=300 K the Li$_{36}$F$_{36}$ structure is 
preserved, although some small distortions in the cage occur.
According to our simulations, the Li$_{36}$F$_{36}$ nanocage 
disintegrates above 600 K, so that in principle this system could be indeed 
produced and experimentally observed at room temperature. 
%In the other ionic nanocages disintegration occurs at ??? K, ??? K,
%and ??? K, for Na$_{36}$F$_{36}$, Li$_{36}$Cl$_{36}$, Na$_{36}$Cl$_{36}$,
%and K$_{36}$Br$_{36}$, respectively, so that... ?

In Table I we also report the percent ionic character, which 
can be evaluated taking into account the electronegativities of the 
constituent atoms and using the Pauling's relation.
As expected C$_{60}$, C$_{72}$, and Si$_{60}$ are characterized
by purely covalent bondings, Li$_{36}$F$_{36}$, Na$_{36}$F$_{36}$,
Li$_{36}$Cl$_{36}$, Na$_{36}$Cl$_{36}$, and K$_{36}$Br$_{36}$ are systems with
a predominant ionic character, while B$_{36}$N$_{36}$ and 
Be$_{36}$O$_{36}$ are partially ionic complexes (B is less electronegative
than N and Be less electronegative than O).
This aspect will be relevant for future considerations.

\begin{table}
\vfill
\eject
\caption{Diameter, cohesive energy $E_c$, HOMO-LUMO energy gap $E_g$, and 
percent ionic character of the bondings in the considered nanocages}
\begin{center}
\begin{tabular}{|c|c|c|c|c|}
\hline
system & diameter (A) & $E_c$ (eV) & $E_g$ (eV) & ionic char. \\ \tableline
\hline
 C$_{60}$          & 7.10 & -7.92 & 1.65 &  0\% \\  
 C$_{72}$          & 8.58 & -7.77 & 1.53 &  0\% \\ 
 B$_{36}$N$_{36}$  & 8.70 & -7.65 & 4.48 & 22\% \\
 Be$_{36}$O$_{36}$ & 8.03 & -6.09 & 4.89 & 58\% \\
 Li$_{36}$F$_{36}$ & 9.36 & -4.16 & 5.86 & 89\% \\
 Si$_{60}$         &11.67 & -3.86 & 0.39 &  0\% \\  
 Na$_{36}$F$_{36}$ &11.20 & -3.53 & 4.79 & 90\% \\ 
 Li$_{36}$Cl$_{36}$&11.30 & -3.18 & 5.68 & 70\% \\  
 Na$_{36}$Cl$_{36}$&13.50 & -2.77 & 4.73 & 71\% \\ 
 K$_{36}$Br$_{36}$ &18.50 & -2.54 & 4.04 & 63\% \\
\hline
\end{tabular}                                                
\end{center}
\label{table1}                                  
\end{table}

The electronic dipole moment of the systems is reported in Tables II-IV,
and is computed as:

\begin{equation}
\mu = -e\int d{\bf r}\, {\bf r}\, n({\bf r}) + \sum_i^{N} Z_i {\bf R}_i \;,
\end{equation}
where $-e$ is the electron charge, $n({\bf r})$ the electronic
number density, and $Z_i$ and ${\bf R}_i$ are the valence and spatial coordinate
of the $i$-th ion of the system, respectively.
All the considered nanocages, in their optimized, isolated structure, are 
characterized by a negligible total dipole moment (in all cases not larger
than 0.2 D). The scenario changes when a small molecule with a finite electronic 
dipole moment is encapsulated into the cages. 
We denote these endohedral complexes as X@Y, where X=HF, LiF, NaCl, and H$_2$O, 
while Y=C$_{60}$, C$_{72}$, B$_{36}$N$_{36}$,
Be$_{36}$O$_{36}$, Li$_{36}$F$_{36}$, Na$_{36}$F$_{36}$, Li$_{36}$Cl$_{36}$,
Na$_{36}$Cl$_{36}$, and K$_{36}$Br$_{36}$ (only the Na$_{36}$Cl$_{36}$ and
K$_{36}$Br$_{36}$ nanocages are sufficiently large
to encapsulate a NaCl molecule without any significant distortion).
When the small molecules are encapsulated into C$_{60}$ or C$_{72}$,
as a consequence of the counteracting dipole moment induced in the
cage, the effective dipole moment of the complex is severely reduced to
a value that is less than 30\% of the dipole moment of the isolated 
molecule. For H$_2$O@C$_{60}$ this confirms our previous findings\cite{Ensing} 
obtained computing the dipole moment using the Wannier-function 
approach\cite{Wannier}
and agrees with experimental dielectric measurements performed at low temperature
and infra-red spectra of H$_2$O@C$_{60}$ obtained at liquid Helium temperature,
which measured a dipole moment of 0.5 $\pm$ 0.1 D\cite{Meier,Shugai}.  
Our calculation is also consistent with the experimental estimate 
(0.45 $\pm$ 0.05 D) reported for HF@C$_{60}$\cite{Krachmalnicoff}
and with previous theoretical estimates for this system\cite{Dolgonos2014}.
Therefore carbon fullerene cages shield more than 70\% of the dipole moment
of the encapsulated molecules and clearly act as molecular Faraday cages.
With the B$_{36}$N$_{36}$ cage the screening effect is less pronounced,
since the dipole moment reduction is about 40\%, while in the case of 
Be$_{36}$O$_{36}$ this reduction amounts to only 10\%,
so that this cage turns out to be
almost  electrically ``transparent'' with X@Be$_{36}$O$_{36}$ that acquires
essentially the same dipole moment of the encapsulated X molecule. 
A {\it qualitatively} different behavior occurs with encapsulation into 
ionic nanocages: in fact an ``anti-screening effect'' is observed since
the dipole moment of the endohedral complexes is significantly 
{\it increased}, by an amount ranging from 20 to 70\%, thus 
indicating that ionic nanocages actually act as dipole-field amplifiers. 
Interestingly, the precise amount of percentage increase depends more on the
specific ionic nanocage than on the dipole moment of the encapsulated
molecule, even considering that LiF and NaCl are characterized by a value 
of the dipole moment much larger than that of HF or H$_2$O.
In particular, ionic nanocages with Na and K atoms exhibit a
more pronounced increase of the dipole moment than those with Li atoms.  
We have verified that 
the dipole moment of the different endohedral complexes does not 
change significantly by replacing the rVV10 DFT functional by the PBE 
one\cite{PBE}, that is a popular functional unable to properly take vdW 
interactions into account,
thus showing that vdW effects are not relevant for this quantity which
is evidently mostly determined by electrostatic interactions.

\begin{table}
\vfill
\eject
\caption{Electronic dipole moment $\mu$ of endohedral complexes with HF molecule inside;
$\delta\mu$ denotes the change of the dipole moment of the endohedral complex 
with respect to that (1.78 D) of the encapsulated HF molecule when it is isolated. 
Binding energy (in square parenthesis using the PBE functional
in place of rVV10) of endohedral complexes.} 
\begin{center}
\begin{tabular}{|c|c|c|c|}
\hline
system & $\mu$ (D) & $\delta\mu$ (D) &  $E_{\rm bind}$ (meV) \\ \tableline
\hline
 HF@C$_{60}$              &  0.52   & -1.26 (-71\%) &  -479 [-64] \\
 HF@C$_{72}$              &  0.47   & -1.31 (-74\%) &  -617 [-61] \\
 HF@B$_{36}$N$_{36}$      &  1.09   & -0.69 (-39\%) &  -374 [-63] \\
 HF@Be$_{36}$O$_{36}$     &  1.63   & -0.15  (-8\%) &  -322 [-97] \\
 HF@Li$_{36}$F$_{36}$     &  2.13   & +0.35 (+20\%) &  -100 [-59] \\
 HF@Na$_{36}$F$_{36}$     &  2.55   & +0.77 (+43\%) &   -52 [-25] \\
 HF@Li$_{36}$Cl$_{36}$    &  2.21   & +0.43 (+24\%) &   -74 [-17] \\
 HF@Na$_{36}$Cl$_{36}$    &  2.58   & +0.80 (+45\%) &   -58  [-9] \\
\hline
\end{tabular}                                                
\end{center}
\label{table2}                                  
\end{table}

\begin{table}
\vfill
\eject
\caption{Electronic dipole moment $\mu$ of endohedral complexes with LiF 
and NaCl molecule inside;
$\delta\mu$ denotes the change of the dipole moment of the endohedral complex 
with respect to those (6.17 D for LiF and 8.59 D for NaCl) of the encapsulated molecules 
when they are isolated. 
Binding energy (in square parenthesis using the PBE functional
in place of rVV10) of endohedral complexes.} 
\begin{center}
\begin{tabular}{|c|c|c|c|}
\hline
system & $\mu$ (D) & $\delta\mu$ (D) &  $E_{\rm bind}$ (meV) \\ \tableline
\hline
 LiF@C$_{60}$             &  1.73   & -4.44 (-72\%) &-1098 [-585] \\
 LiF@C$_{72}$             &  1.62   & -4.55 (-74\%) &-1088 [-538] \\
 LiF@B$_{36}$NF$_{36}$    &  3.80   & -2.37 (-38\%) & -958 [-541] \\
 LiF@Be$_{36}$O$_{36}$    &  5.52   & -0.65 (-11\%) & -877 [-569] \\
 LiF@Li$_{36}$F$_{36}$    &  7.98   & +1.81 (+29\%) & -402 [-314] \\
 LiF@Na$_{36}$F$_{36}$    &  7.99   & +1.82 (+29\%) & -215 [-175] \\
 LiF@Li$_{36}$Cl$_{36}$   &  7.55   & +1.38 (+22\%) & -225 [-150] \\
 LiF@Na$_{36}$Cl$_{36}$   &  8.63   & +2.46 (+40\%) & -158  [-95] \\
\hline
 NaCl@Na$_{36}$Cl$_{36}$  & 12.1    & +3.51 (+41\%) & -325 [-200] \\
 NaCl@K$_{36}$Br$_{36}$   & 14.9    & +6.31 (+73\%) & -170 [-110] \\
\hline
\end{tabular}                                                
\end{center}
\label{table3}                                  
\end{table}

\begin{table}
\vfill
\eject
\caption{Electronic dipole moment $\mu$ of endohedral complexes with H$_2$O 
molecule inside;
$\delta\mu$ denotes the change of the dipole moment of the endohedral complex 
with respect to that (1.86 D) of the encapsulated H$_2$O molecule when it is isolated. 
Binding energy (in square parenthesis using the PBE functional
in place of rVV10) of endohedral complexes.} 
\begin{center}
\begin{tabular}{|c|c|c|c|}
\hline
system & $\mu$ (D) & $\delta\mu$ (D) &  $E_{\rm bind}$ (meV) \\ \tableline
\hline
 H$_2$O@C$_{60}$          &  0.52   & -1.34 (-72\%) & -554   [+9] \\
 H$_2$O@C$_{72}$          &  0.52   & -1.34 (-72\%) & -531  [-25] \\
 H$_2$O@B$_{36}$N$_{36}$  &  1.17   & -0.69 (-37\%) & -547  [-84] \\
 H$_2$O@Be$_{36}$O$_{36}$ &  1.61   & -0.25 (-13\%) & -535 [-187] \\
 H$_2$O@Li$_{36}$F$_{36}$ &  2.20   & +0.34 (+18\%) & -152  [-76] \\
 H$_2$O@Na$_{36}$F$_{36}$ &  2.68   & +0.82 (+44\%) &  -73  [-36] \\
 H$_2$O@Li$_{36}$Cl$_{36}$&  2.32   & +0.46 (+25\%) &  -87  [-25] \\
 H$_2$O@Na$_{36}$Cl$_{36}$&  2.81   & +0.95 (+51\%) &  -47  [-13] \\
\hline
\end{tabular}                                                
\end{center}
\label{table4}                                  
\end{table}

In order to better elucidate the mechanisms underlying the 
dipole-moment variations, a further analysis is performed. 
In Fig. 4 the changes in electron distribution, resulting
from the encapsulation process are shown, for the HF@C$_{72}$ 
endohedral complex by plotting
the {\it differential} 
charge density, $\Delta \rho$, defined as the difference between the total 
electron density of the whole system and the superposition of the densities of 
the separated fragments (HF molecule and C$_{72}$ cage), keeping the 
same geometrical structure and atomic positions that these
fragments have within the whole optimized system.
This procedure is justified since, for non-ionic nanocages,
the changes of atomic positions upon encapsulation of a small
molecule are very small.
Note that, in line with previous observations\cite{Dolgonos2014}, the polar 
molecule HF is slightly displaced from the center
of the nanocages, so that the H-cage distance is smaller than the 
F-cage one, due to the attractive electrostatic interactions 
between the H atom and the atoms of the cages.
In Fig. 5 we also plot the one-dimensional profile $\Delta \rho (z)$,
computed along the F-H $z$ axis, as a function of
$z$ values, by integrating $\Delta \rho$ over the corresponding,
orthogonal $x,y$ planes.  
Inspection of these figures, that are representative of what happens in 
all the nanocages where
a significant screening effect is observed, reveals that in HF@C$_{72}$
there is a pronounced electron charge accumulation in the region between the
H atom and the cage with a charge depletion around the F atom, leading to the
formation of the counteracting dipole moment which considerably reduces the
effective dipole moment of the endohedral complexes; 
clearly the 
overall response of these nanocages to the HF molecule dipole moment is a 
significant charge density shift.
One can make quantitative the information contained in Fig. 5 
by evaluating the induced dipole moment as: 

\begin{equation}
\mu_{ind} = -\int dz\, z\, \Delta \rho (z) \;,
\end{equation}
where $\Delta \rho (z)$ has been defined above.
The numerical value of $\mu_{ind}$ is found to essentially coincide (see Fig. 5) with
that of $\delta\mu$ reported in Table II for HF@C$_{72}$.

Instead, in predominantly ionic nano-cages, the dipole moment of the 
endohedral complexes is {\it increased} from the value of the isolated molecule 
that is encapsulated ({\it anti-screening effect}) 
and this dipole amplifications is mostly due to a slight cage distortion:
in fact positive ions (Li+, Na+, and K+) are displaced with respect to the 
negative ions (F-, Cl-, and Br-). We can quantify this effect by computing the
distance between the position of the center of mass of the positive ions
and that of the center of mass of the negative ions: this quantity
ranges from $3.3 \times 10^{-3}$ \AA\ for HF@Li$_{36}$F$_{36}$ to 
$5.5 \times 10^{-2}$ \AA\ for NaCl@K$_{36}$Br$_{36}$. 
Basically, as a small molecule with a permanent dipole moment 
is encapsulated into a ionic nanocage this reacts in such a way
to displace the positive ions with respect to the negative ions
along the direction of the molecule dipole moment.
This cage distortion accounts for more than 80\% of the observed
increase of the dipole moment; the remaining increase is due to
the electronic charge polarization, and the small change of the interatomic 
distance of the encapsulated diatomic molecule.
The basic mechanism of dipole increase is illustrated in Fig. 6 for 
NaCl@Na$_{36}$Cl$_{36}$, where this effect is pronounced: as can be seen,
the Na+ ions of the Na$_{36}$Cl$_{36}$ nanocage undergo a significant positive
displacement below the Cl atom of the encapsulated Na-Cl molecule,
while the Cl- ions undergo a negative displacement particularly above the
Na atom of the molecule.
Also in partially-ionic nanocages, B$_{36}$N$_{36}$ and Be$_{36}$O$_{36}$,
upon encapsulation one can detect a relative displacement of the center of
mass of the two kinds of atoms, however this displacement is much smaller
(by one or two orders of magnitude) 
than that observed in predominantly ionic nano-cages, so that the more standard
mechanism of screening prevails.

\begin{figure}
%{\vskip 1.3cm}
\centerline{
\includegraphics[width=13cm]{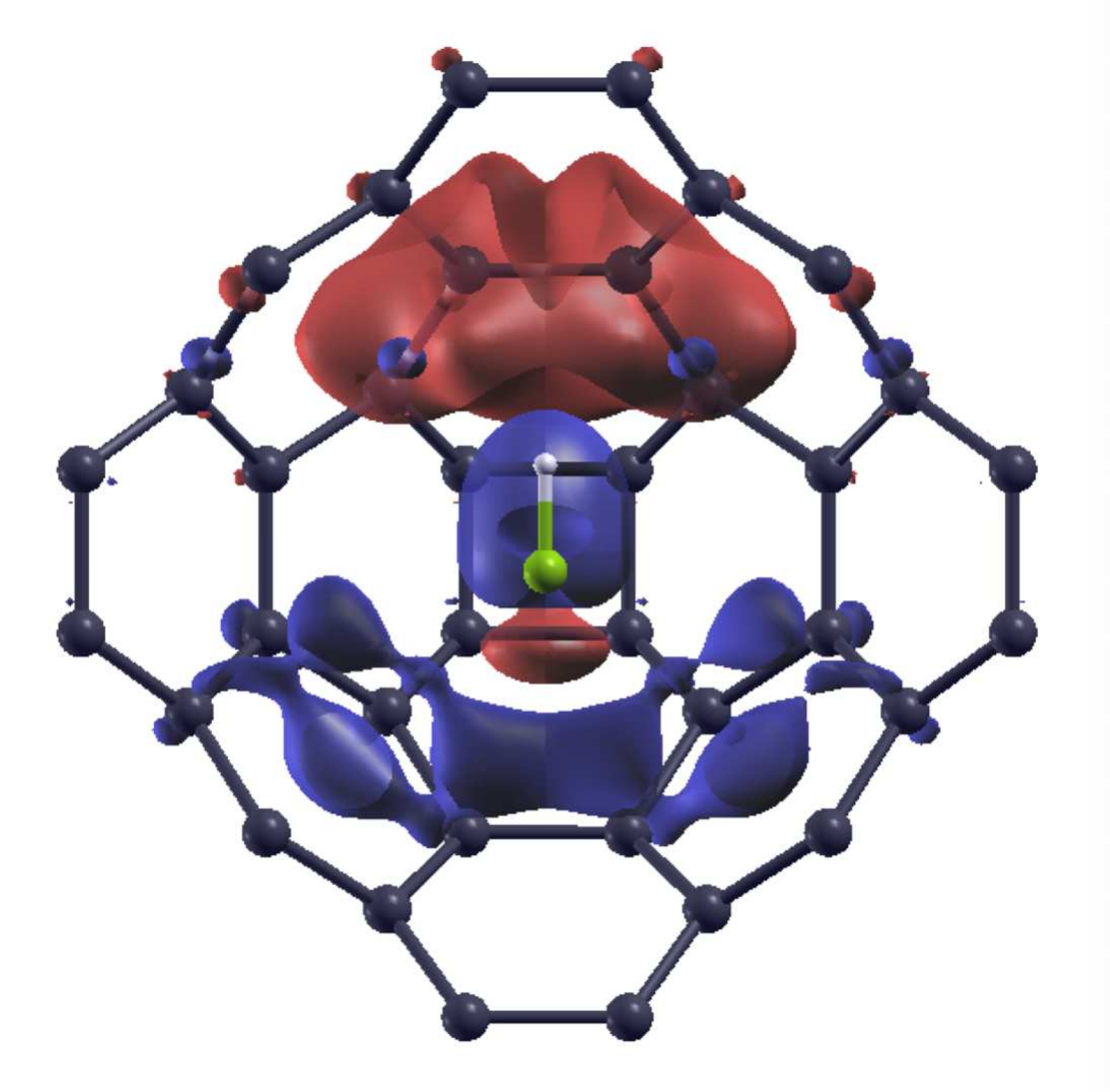}
}
\caption{Differential electron charge density, $\Delta \rho$, 
for HF@C$_{72}$, with isosurfaces shown at 
$\pm 2\times 10^{-3}\, e/{\rm \AA}^3$.
Red areas indicate electron density gain, while blue areas indicate loss
of electron density relative to the empty C$_{72}$ cage and the 
isolated HF molecule.} 
\label{fig4}
%\huge
\end{figure}
\eject
                      
\begin{figure}
%{\vskip 1.3cm}
\centerline{
\includegraphics[width=16cm,angle=-90]{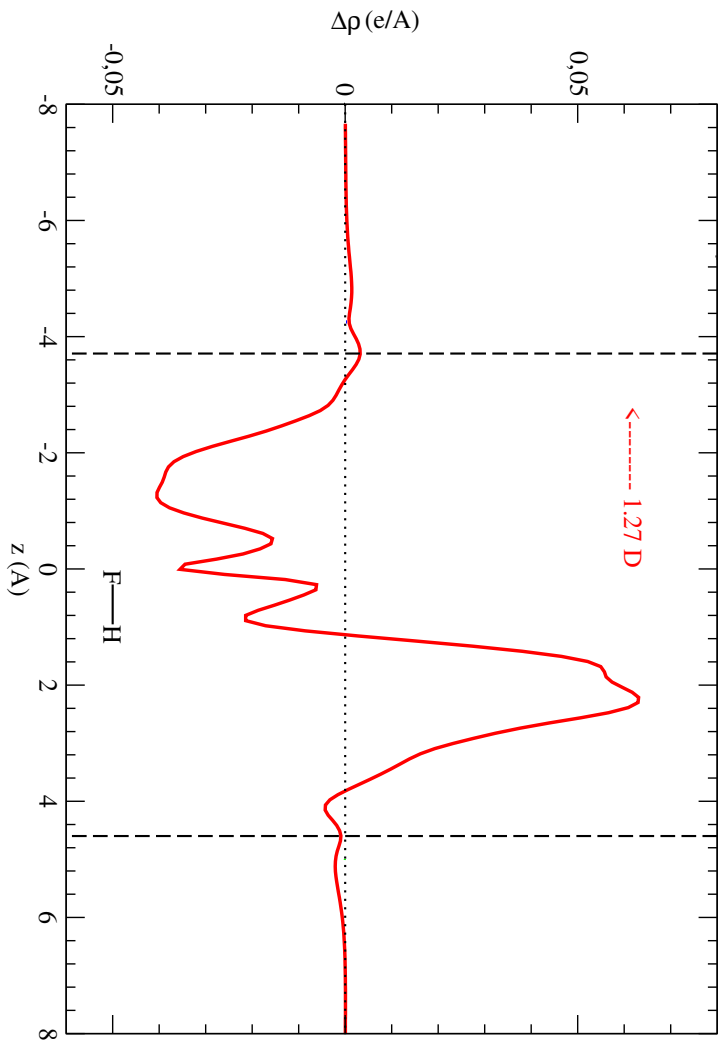}
}
\caption{Differential electron charge density, $\Delta \rho (z)$, along the
F-H $z$ axis, in HF@C$_{72}$.
The vertical, black, dotted lines indicate the positions of the cage surface, 
while the red arrow represents the induced dipole moment with a numerical
value obtained by integration on the $z$ axis (see text).}   
\label{fig5}
%\huge
\end{figure}
\eject

\begin{figure}
%{\vskip 1.3cm}
\centerline{
\includegraphics[width=16cm,angle=0]{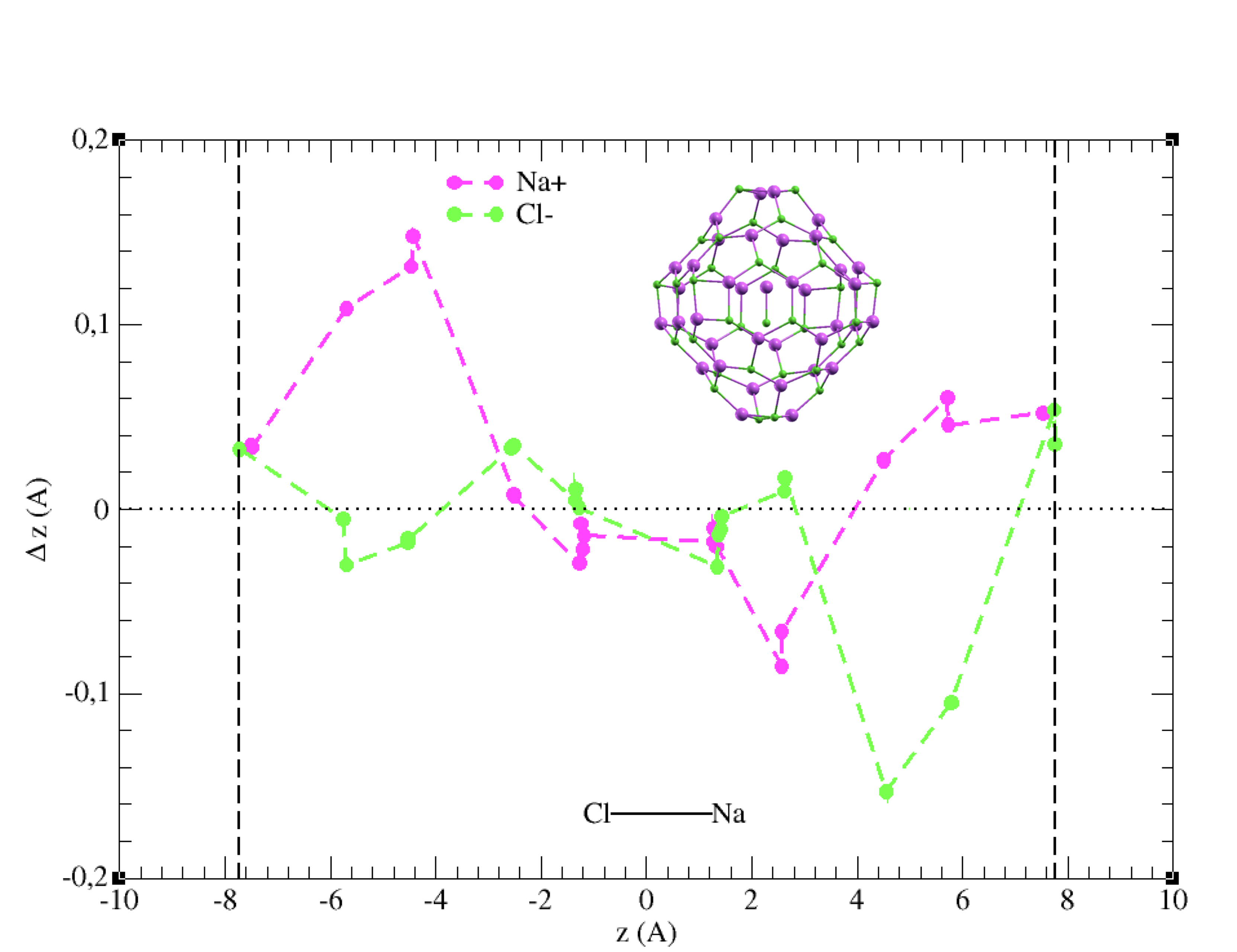}
}
\caption{NaCl@Na$_{36}$Cl$_{36}$ complex: displacements,
along the Cl-Na $z$ axis, of Na+ and Cl- ions of the
Na$_{36}$Cl$_{36}$ ionic nanocage upon encapsulation of NaCl: small circles
represent the actual displacements of the ions, while the dashed lines
are just a guide for the eye.
The vertical, black, dotted lines indicate the positions of the cage surface.} 
\label{fig6}
%\huge
\end{figure}
\eject

In Table II we also report the binding energies of endohedral complexes;
this is computed as the difference between the total energy
of the X@Y complex and the sum of the total energies of the constituent
parts X and Y.
We also add the binding energies obtained by replacing the rVV10 DFT 
functional with the PBE one.
As can be seen, all the molecules are found to form stable complexes with the 
considered nanocages (with the exception of H$_2$O@C$_{60}$ using the PBE functional); 
however, differently from what found for the dipole moment, a proper inclusion of 
vdW effects is here crucial since these
account for the dominant part of the binding energy between the cage and the
HF and H$_2$O molecules, and represent a significant contribution also for the binding
with the LiF and NaCl molecules, where electrostatic 
and induction-polarization interactions are important, as shown
by the fact that the binding energy predicted by the PBE functional
is (in absolute value) not much smaller that that obtained by the 
vdW-corrected rVV10 functional. 
Note that in the present systems zero-point energy effects are expected to 
be small\cite{Dolgonos2014}. 

To complement the above first-principles DFT analysis, it is instructive to perform 
simplified SCS calculations for the dipole screening of a few relevant complexes. 
Within SCS the inner molecule is treated as a pointlike dipole, so that some information
about the actual molecule-fullerene interaction is lost. Moreover, given the linearity of 
the approach, screening is independent from the magnitude of the molecular dipole. We will 
thus consider encapsulated HF as the reference geometry, in order to extract 
qualitative trends.
We note that the DFT dipole screening for C$_{72}$ is qualitatively
reproduced by SCS, reinforcing our previous conclusions: a reduction of the molecular 
dipole by 57\% is found, although
SCS cannot account for the quasi-metallic orbitals responsible for the charge localization 
observed in Fig.~\ref{fig4}.  In fact, while the charge of each Drude oscillator cannot 
move much from its initial position, the many-body coupling between all oscillators 
produces larger displacements where also DFT charge rearrangements are more concentrated.
This result is compatible with the long-ranged charge oscillations predicted in 
low-dimensional nanostructures~\cite{science} by a related Drude model: many-body couplings 
can strongly enhance the non-locality~\cite{jpcl} of the density response
in low dimensionality.

On the other hand, SCS, differently from what obtained by DFT, 
is found to roughly reproduce the same screening mechanism in 
all fullerene cages.
For instance, the SCS screening is overestimated in B$_{36}$N$_{36}$, where a
dipole reduction by 55\% is found.
Given the limitations of the naive Tkatchenko-Scheffler method
in describing polar materials, we exploited here ionic polarizability data taken from
ref.\cite{gould}, as prescribed.
Even more interesting, in Li$_{36}$F$_{36}$, no antiscreening is observed with SCS: 
the coupled atomic polarizabilities alone are clearly insufficient to cause dipole 
increase (in fact, even in this system SCS predicts a small screening of about 4\%). 
This reconfirms that the leading antiscreening mechanism in ionic fullerenes is 
indeed played by the ionic rearrangements, that cannot be captured by SCS. 
We observe here the analogy with ionic crystals, where strong electron-phonon coupling 
can be associated to the rise of polarons~\cite{polaron}, which involve a major interplay 
between charge localization and structural distortions.

\section{Conclusions}
We have presented the results of a first-principles study of 
screening effects in endohedral complexes made by small molecules,
with a finite electronic dipole moment, encapsulated into different 
nanoscale cages. A detailed analysis of the effective dipole moment of the
complexes and of the electronic charge distribution suggests that
screening effects crucially depend on the nature of the
intramolecular bonds of the cage: screening is maximum
in covalent-bond carbon nanocages, while it is reduced in partially-ionic
nanocages B$_{36}$N$_{36}$ and Be$_{36}$O$_{36}$, being very small in the
latter cage which turns out to be almost ``electrically transparent''. 
Interestingly, in the case of the ionic-bond nanocages, an 
{\it antiscreening} effect is observed: in fact, due to the relative displacement 
of positive and negative ions, induced by the dipole moment of the 
encapsulated molecule, these cages act as dipole-field amplifiers.
Our results open the way to the possibility of tuning the dipole moment of
nanocages and of generating electrostatic fields at the nanoscale without the 
aid of external potentials. Moreover, we can expect some transferability of the 
observed screening effects also to nanotubes and 2D materials.

\section{Acknowledgements}
We acknowledge funding from Fondazione Cariparo, Progetti di Eccellenza 2017, 
relative to the project “Engineering van der Waals Interactions: Innovative 
paradigm for the control of Nanoscale Phenomena”.

\vfill
\eject

\end{document}